\documentclass[%
 reprint,
 superscriptaddress,
 amsmath,amssymb,
 aps,
]{revtex4-2}

\usepackage{graphicx}
\usepackage{dcolumn}
\usepackage{bm}

\usepackage{physics}
\usepackage{xcolor}
\usepackage{mathtools}
\usepackage{amssymb}
\usepackage{color,soul}
\usepackage{float}

\begin{document}


\title{Enhanced strong interaction effect in synthetic spin-orbit coupling with mixed spin symmetry}

\author{Ayaka Usui}
\email{ayaka.usui@uab.cat}
\affiliation{Departament de F\'{i}sica Qu\`{a}ntica i Astrof\'{i}sica, Universitat de Barcelona, Mart\'{i} i Franqu\'{e}s, 1, E08028 Barcelona, Spain}
\affiliation{Institut de Ci\`{e}ncies del Cosmos (ICCUB), Universitat de Barcelona, Mart\'{i} i Franqu\'{e}s, 1, E08028 Barcelona, Spain}
\affiliation{Departament de F\'{i}sica, Universitat Aut\`{o}noma de Barcelona, 08193 Bellaterra, Spain}

\author{Abel Rojo-Franc\`{a}s}
\affiliation{Departament de F\'{i}sica Qu\`{a}ntica i Astrof\'{i}sica, Universitat de Barcelona, Mart\'{i} i Franqu\'{e}s, 1, E08028 Barcelona, Spain}
\affiliation{Institut de Ci\`{e}ncies del Cosmos (ICCUB), Universitat de Barcelona, Mart\'{i} i Franqu\'{e}s, 1, E08028 Barcelona, Spain}

\author{James Schloss}
\affiliation{Massachusetts Institute of Technology, Cambridge, MA, United States}

\author{Bruno Juli\'{a}-D\'{i}az}
\affiliation{Departament de F\'{i}sica Qu\`{a}ntica i Astrof\'{i}sica, Universitat de Barcelona, Mart\'{i} i Franqu\'{e}s, 1, E08028 Barcelona, Spain}
\affiliation{Institut de Ci\`{e}ncies del Cosmos (ICCUB), Universitat de Barcelona, Mart\'{i} i Franqu\'{e}s, 1, E08028 Barcelona, Spain}


\begin{abstract}

Synthetic spin-orbit coupling in cold atoms couples the pseudo-spin and spatial degrees of freedom, and therefore the inherent spin symmetry of the system plays an important role. In systems of two pseudo-spin degrees, two particles contain symmetric states and anti-symmetric states, but the spin symmetry can be mixed for more particles.
This mixed spin symmetry has been overlooked and has not been investigated thoroughly.
We study the role of mixed spin symmetry in the presence of spin-orbit coupling and consider the system of three bosons with two hyper-fine states trapped in a harmonic potential. We investigate the ground state and the energy spectrum by implementing exact diagonalization. 
Similarly to two-particle systems, the interplay between spin-orbit coupling and repulsive interactions between anti-aligned pseudo-spins increases the population of the unaligned spin components in the ground state. Also, the emergence of the mixed spin symmetric states compensates for the rise of the interaction energy.
In contrast to two-particle systems, the pair correlation of the ground state is analogous to the Tonks-Girardeau gas even for relatively small contact interactions, and this feature is enhanced by the spin-orbit coupling.

\end{abstract}

\maketitle


\section{Introduction}

Spin-orbit coupling (SOC) was originally discussed in the system of charged particles and, for instance, studied in the context of the spin Hall effect~\cite{Nagaosa2010Anomalous} or topological insulators states~\cite{Hasan2010Topological}. 
However, by using cold atomic systems it is possible to create synthetic SOC in neutral (pseudo) spin-1/2 bosons~\cite{Lin2011Spin,Zhang2012Collective}, spin-1 Bose gases~\cite{Campbell2016Magnetic} and also in Fermi gases~\cite{Wang2012Spin,Cheuk2012Spin}. In contrast to condensed matter systems, cold atoms can provide tunable and clean platforms and allow us to explore all possible states generated by SOC. 

While SOC in cold atoms is often discussed in the context of the mean-field regime, e.g.~\cite{Li2012Quantum,Zhang2012Mean,Zhang2016Properties}, the regime beyond it has also been attracting attention, such as the regime of interactions that are strong enough to observe entanglement generation~\cite{Sorensen2001Many,Pezze2018Quantum,Wilson2022Beyond}. To bridge single-particle physics and many-body physics in SOC systems, a mapping between cold atomic SOC systems and the Dicke model has been proposed~\cite{Hamner2014Dicke}. So far, the results have only been experimentally demonstrated in the mean-field regime. The work hinted that this mapping worked even beyond the mean-field regime, but the validity of the mapping is not obvious. The assumption imposed for the mapping is that all particles occupy the same real space state, which automatically leads to the pseudo-spin state spanning symmetric spin space only. This assumption is commonly used for Bose gases with two internal degrees of freedom~\cite{Sorensen2001Many}, however SOC couples pseudo-spin states with real space states and allows pseudo-spin states to get out of symmetric spin space. 
Overlooking this contradiction, it was predicted by Refs.~\cite{Lian2013Orbit,Huang2015Spin} that spin squeezing is observed in the mean-field regime in the SOC system by applying this mapping. This essentially would mean that entanglement is generated in a non-interacting system, which is not physical.



Motivated by this contradiction, we tackle the problem of multi-particle systems with SOC by computing low energy states exactly. 
The two-particle system with SOC in one dimension has already been investigated by performing exact diagonalization, and the competition between SOC and contact interactions has also been studied~\cite{Usui2020Spin}. It has been shown that the ground state includes the anti-symmetric pseudo-spin state, which does not feel contact interactions.
However, these results cannot be extended to many-particle systems straightforwardly, because they are rooted in the spin symmetry of only two particles. While two pseudo-spin states are classified into either symmetric or anti-symmetric pseudo-spin states, the spin symmetry can be mixed in more-particle systems, and it is unclear whether the properties of the ground state in the two-particle system are inherited in more-particle systems.
Moreover, pseudo-spin states that are not symmetric have not had as much attention as symmetric pseudo-spin states and have not been investigated as thoroughly. 




In this work, we investigate the role of mixed spin symmetry in SOC systems. To this end, we consider the smallest system with mixed spin symmetry, a three-particle system with two internal states, and we focus on the comparison with two-particle systems. We have built the Hamiltonian of three bosons with two internal states in the presence of SOC by considering the second quantisation and have computed a few low-energy states.
In comparison to two particle systems, we have found similarity in the appearance of mixed symmetric pseudo-spin states in the ground state, which is confirmed for strong interaction between anti-aligned pseudo-spins and reduced interaction energy.
This is also seen in two-particle systems when the anti-symmetric pseudo-spin states emerge.
Also, the emergence of non-symmetric pseudo-spin states challenges the mapping proposed in Ref.~\cite{Hamner2014Dicke}.
At the same time, we have also studied the spatial structure of the ground state by looking at the pair correlation and find that SOC assists contact interaction in inducing a strong interaction effect. Even for relatively small contact interactions, the pair correlation of the ground state has a structure similar to the strongly correlated Tonks-Girardeau gas. 
The pair correlation of the two-particle system is simply the total density distribution and does not display such a strong interaction effect.
It is worth noting that the strong interaction effect enhanced by SOC is analogous to Ref.~\cite{Chen2020Spin}, which shows that SOC amplifies the spin squeezing.


The structure of this paper is as follows. 
First, in Sec.~\ref{sec:symmetry} we review the spin symmetry in pseudo-spin-1/2 systems and clarify the existence of non-symmetric pseudo-spin states.
Section~\ref{sec:basis} explains the formulation of our system. We introduce the Hamiltonian of our system in the first quantization representation and then transform it into the second quantization representation. This is necessary to implement exact diagonalization. Section~\ref{sec:groundstate} shows our numerical results of the ground state and the energy spectrum. We consider two cases, (i) repulsive interaction between anti-aligned pseudo-spins and (ii) repulsive interaction between aligned pseudo-spins. The conclusion is given in Sec.~\ref{sec:conclusions}. 


\section{Spin symmetry of bosons with two internal states} \label{sec:symmetry}

Before the main part of this work, we will review the symmetry of pseudo-spin-1/2 systems and clarify what mixed spin symmetry means in our context.  
First, let us start with two-particle states. Since we are considering bosons, every state is symmetric under permutation. The considered system has pseudo-spin and spatial degrees of freedom, and the pseudo-spin basis are classified into three symmetric pseudo-spin states, $\ket{\downarrow\downarrow}$, $(\ket{\downarrow\uparrow}+\ket{\uparrow\downarrow})/\sqrt{2}$, $\ket{\uparrow\uparrow}$ and one anti-symmetric pseudo-spin state $(\ket{\downarrow\uparrow}-\ket{\uparrow\downarrow})/\sqrt{2}$. To satisfy the bosonic symmetry, the spatial states of symmetric pseudo-spin states are symmetric, i.e. even functions in the relative coordinate $x_1-x_2$. Similarly, the spatial states of anti-symmetric pseudo-spin states are anti-symmetric, i.e. odd functions in the relative coordinate $x_1-x_2$. Since the contact interaction occurs at a point and is described as the delta function $\delta(x_1-x_2)$, the anti-symmetric spatial states do not feel the contact interaction. 

By adding more particles to the system, the spin symmetry is mixed, and the anti-symmetry under all permutation does not exist anymore. 
Specifically, states of three bosons with pseudo-spin-1/2 can be constructed with the following symmetric spin basis, 
\begin{align} \label{eq:basis_symmetric}
    \ket{S_1}
    &=\ket{\downarrow\downarrow\downarrow}
    \nonumber\\
    \ket{S_2}
    &=\frac{1}{\sqrt{3}}\left(
    \ket{\downarrow\downarrow\uparrow} + \ket{\downarrow\uparrow\downarrow} + \ket{\uparrow\downarrow\downarrow}
    \right)
    \nonumber\\
    \ket{S_3}
    &=\frac{1}{\sqrt{3}}\left(
    \ket{\downarrow\uparrow\uparrow} + \ket{\uparrow\downarrow\uparrow} + \ket{\uparrow\uparrow\downarrow}
    \right)
    \nonumber\\
    \ket{S_4}
    &=\ket{\uparrow\uparrow\uparrow},
\end{align}
and the following mixed spin symmetric basis,
\begin{align} \label{eq:basis_mixspin}
    \ket{M_1}
    &=
    \frac{-1}{\sqrt{6}}\left(
    \ket{\downarrow\uparrow\uparrow}+\ket{\uparrow\downarrow\uparrow}-2\ket{\uparrow\uparrow\downarrow}
    \right)
    \nonumber\\
    \ket{M_2}
    &=\frac{1}{\sqrt{6}}\left(
    \ket{\uparrow\downarrow\downarrow}+\ket{\downarrow\uparrow\downarrow}-2\ket{\downarrow\downarrow\uparrow}
    \right)
    \nonumber\\
    \ket{M_3}
    &=\frac{1}{\sqrt{2}}\left(
    \ket{\uparrow\downarrow\uparrow} - \ket{\downarrow\uparrow\uparrow}
    \right)
    \nonumber\\
    \ket{M_4}
    &=\frac{1}{\sqrt{2}}\left(
    \ket{\uparrow\downarrow\downarrow} - \ket{\downarrow\uparrow\downarrow}
    \right)
    .
\end{align}
For example, switching the first spin and the second spin in the pseudo-spin basis $\ket{M_1}$ changes nothing (symmetric), but by switching the second spin and the third spin, the resultant state can no longer be described with $\ket{M_1}$. Therefore, these bases are not symmetric or anti-symmetric, but are instead mixed symmetric. 
To satisfy symmetry under all permutations, 
the spatial states of the mixed spin symmetric states are formed such that the total state is symmetric.


When spatial states and the pseudo-spin states are decoupled in systems of pseudo-spin-1/2 bosons, the pseudo-spin space is confined to the symmetric spin space. For instance, a pseudo-spin-1/2 Bose-Einstein condensate in a harmonic trap can be modelled with the collective spin operators, and the pseudo-spin Hamiltonian is expressed as $\hat{H}=g\hat{S}_z^2$~\cite{Sorensen2001Many}. The influence of the spatial state is included in the interaction strength $g$, and the strength $g$ can be tuned by modifying the potential for example. 
However, as we introduce in the next section, SOC couples the spatial degree and the pseudo-spin degree.
Thus, non-symmetric spin states play as an important role as symmetric spin states in the system with SOC. 

\section{Few interacting bosons with SOC} \label{sec:basis}

We present our formalism here. 
First, we introduce the Hamiltonian of our system in the first quantization basis and then rewrite it in the second quantization basis to induce the bosonic symmetry. This approach is often used to present the Hamiltonian of few particle systems, e.g.~\cite{Mujal2017Quantum,Mujal2020Spin,Rojo2020Static}, and we refer interested readers to a detailed reference~\cite{Raventos2017Cold}.
Although we consider three particles in this work, the approaches explained here work for any particle number, and we keep the discussion general in this section.

\subsection{Hamiltonian in the first quantisation basis}

We consider a few bosons trapped in a one-dimensional harmonic potential with two internal degrees of freedom in the presence of SOC. The Hamiltonian reads
\begin{align} \label{eq:Htotal}
    \hat{H}
    &=
    \hat{H}_0 + \hat{H}_{\mathrm{int}}
    .
\end{align}
The single-particle Hamiltonian is given by 
\begin{align} \label{eq:H0}
    \hat{H}_0
    &=
    \sum_{j=1}^{N}
    \left(
    \frac{\hat{p}_j^2}{2m}
    +
    \frac{m\omega^2 \hat{x}_j^2}{2}
    +
    \frac{\hbar k}{m} \hat{p}_j \hat{\sigma}_z^{(j)}
    +
    \frac{\hbar\Omega}{2}\hat{\sigma}_x^{(j)}
    \right)
    ,
\end{align}
where $N$ is the particle number, $m$ is the mass of each particle, $\omega$ is the harmonic trap frequency, $k$ is the SOC strength, and $\Omega$ is the Raman coupling strength. The $\sigma_{x,z}^{(j)}$ are the Pauli matrices. The third term describes SOC between real space and pseudo-spin space, and the momentum of the $j$th particle is shifted as $\hat{p}_j\pm\hbar k$ for up spin and down spin, respectively. Experimentally, this coupling is induced by the momentum difference between the two Raman laser pulses~\cite{Lin2011Spin}, and $k$ is the projected wave number determined by the wavelength and the angle of intersection between the Raman lasers~\cite{HamnerThesis}. Without either of the third and the fourth terms, the single-particle Hamiltonian can be diagonalised with the basis of $\hat{\sigma}_z$ or $\hat{\sigma}_x$, although the Raman coupling cannot be turned off completely while keeping SOC due to the implementation of SOC.
In this work, we consider repulsive contact interactions where particles repel only when two particles meet. The interaction Hamiltonian can be decomposed into three contributions as
\begin{align} \label{eq:Hint}
    \hat{H}_{\mathrm{int}}
    &=
    \hat{H}_{\downarrow\downarrow} + \hat{H}_{\downarrow\uparrow} + \hat{H}_{\uparrow\uparrow}
    .
\end{align}
Each of the components describes contact interaction and is given by
\begin{subequations} \label{eq:Hint_1}
\begin{align}      
\hat{H}_{\downarrow\downarrow}
=
\sum_{i<j}
g_{\downarrow\downarrow}\delta(x_i-x_j)
\ket{\downarrow}_i \ket{\downarrow}_j 
\bra{\downarrow}_i \bra{\downarrow}_j
,
\end{align}
\begin{align}      
\hat{H}_{\uparrow\uparrow}
=
\sum_{i<j}
g_{\uparrow\uparrow}\delta(x_i-x_j)
\ket{\uparrow}_i \ket{\uparrow}_j 
\bra{\uparrow}_i \bra{\uparrow}_j
,
\end{align}
\begin{align}      
\hat{H}_{\downarrow\uparrow}
&=
\sum_{i<j}
g_{\downarrow\uparrow}\delta(x_i-x_j)
\nonumber\\ 
&\quad\quad
\left(
\ket{\downarrow}_i \ket{\uparrow}_j 
\bra{\downarrow}_i \bra{\uparrow}_j
+
\ket{\uparrow}_i \ket{\downarrow}_j 
\bra{\uparrow}_i \bra{\downarrow}_j
\right),
\end{align}
\end{subequations}
where $g_{\downarrow\downarrow},g_{\uparrow\uparrow},g_{\downarrow\uparrow}$ are the interaction strength.

The above formalism is called the first quantization and labels each particle as seen in Eqs.~\eqref{eq:H0}\eqref{eq:Hint_1}. This is suitable to distinguishable particles although it can work for indistinguishable particles after some modification, for example see Ref.~\cite{Usui2020Spin}.
To express indistinguishable particles simply, we implement the second quantization representation in a truncated space.

\subsection{Hamiltonian in the second quantization basis}\label{sec:2ndquantization}



Consider $N$ particles occupying $N$ of first $M$ eigenstates of the harmonic oscillator for down spins and up spins. Defining the creation and annihilation operators for the $j$th eigenstate as $\hat{a}_{j}$ and $\hat{a}_{j}^{\dagger}$, the second quantised version of the single-particle part~\eqref{eq:H0} is given by 
\begin{align}\label{eq:H0_2nd}
    \hat{H}_0
    &=
    \sum_{i,j=1}^{2M}
    \hat{a}^{\dagger}_i\hat{a}_j
    \epsilon_{i,j}
    ,
\end{align}
where the indices $i,j$ include the label of the eigenstates of the harmonic oscillator and the label of the pseudo-spin degrees. Specifically, we take numbers in $[1,M]$ for the eigenstates for down spins and numbers in $[M+1,2M]$ for the eigenstates for up spins.
The single-particle energy is represented as~\cite{Mujal2020Spin}
\begin{align}
    \epsilon_{i,j}
    &=
    \hbar\omega
    \left(n_x(i) + \frac{1}{2}
    \right)
    \delta_{i,j}
    +
    \frac{i k\ell}{\sqrt{2}}
    m_{\text{s}}(j)
    \delta_{m_{\text{s}}(i),m_{\text{s}}(j)}
    \nonumber\\
    &\quad
    \times
    \left(
    \sqrt{n_x(j)+1}\delta_{n_x(i),n_x(j)+1}
    - 
    \sqrt{n_x(j)}\delta_{n_x(i),n_x(j)-1}
    \right)
    \nonumber\\\
    &\quad+
    \frac{\hbar\Omega}{2}
    \delta_{m_{\mathrm{s}}(i),-m_{\mathrm{s}}(j)}
    \delta_{n_x(i),n_x(j)}
    ,
\end{align}
where $\ell=\sqrt{\hbar/m\omega}$ is the trap length, $n_x(j)$ represents the label of eigenstate of the harmonic oscillator, and $m_{\text{s}}(j)$ represents the pseudo-spin state: down spins give $m_{\text{s}}(j)=-1$, and up spins give $m_{\text{s}}(j)=1$.  
The interaction part~\eqref{eq:Hint} is given by
\begin{align}\label{eq:Hint_2nd}
    \hat{H}_{\text{int}}
    &=
    \frac{1}{2}
    \sum_{i,j,k,l=1}^{2M}
    \hat{a}_i^{\dagger} \hat{a}_j^{\dagger} \hat{a}_k \hat{a}_l
    V_{i,j,k,l}
    \nonumber\\
    &\quad
    \times\bigg(
    g_{\downarrow\downarrow}
    \delta_{m_{\text{s}}(i),-1} \delta_{m_{\text{s}}(j),-1} \delta_{m_{\text{s}}(k),-1} \delta_{m_{\text{s}}(l),-1}
    \nonumber\\
    &\quad\quad
    +
    g_{\uparrow\uparrow}
    \delta_{m_{\text{s}}(i),1} \delta_{m_{\text{s}}(j),1} \delta_{m_{\text{s}}(k),1} \delta_{m_{\text{s}}(l),1}
    \nonumber\\
    &\quad\quad
    +
    g_{\downarrow\uparrow}
    \big(
    \delta_{m_{\text{s}}(i),1} \delta_{m_{\text{s}}(j),-1} \delta_{m_{\text{s}}(k),1} \delta_{m_{\text{s}}(l),-1} 
    \nonumber\\
    &\quad\quad\quad
    + \delta_{m_{\text{s}}(i),-1} \delta_{m_{\text{s}}(j),1} \delta_{m_{\text{s}}(k),-1} \delta_{m_{\text{s}}(l),1}
    \big)
    \bigg)
    ,
\end{align}
where
\begin{align}\label{eq:Vijkl}
    V_{i,j,k,l}
    &=
    \int_{-\infty}^{\infty}
    dx \quad
    \phi_{n_x(i)}(x) 
    \phi_{n_x(j)}(x) 
    \phi_{n_x(k)}(x) 
    \phi_{n_x(l)}(x)
\end{align}
with $\phi_{n}$ the $n$th eigenstate of harmonic oscillator. The above integral~\eqref{eq:Vijkl} can be solved analytically but is hard to compute for large indices as binomial coefficients appear. See Ref.~\cite{Rojo2020Static} for an efficient calculation of it. 

\begin{figure}[t]
\centering
\includegraphics[width=.9\linewidth]{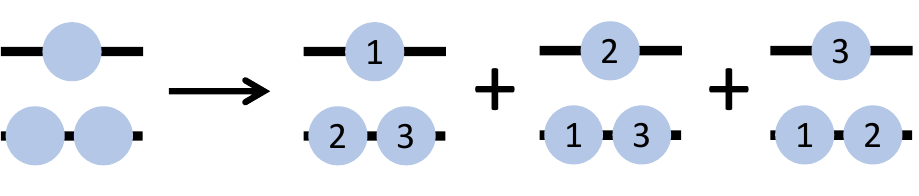}
\caption{Example of transforming the second quantization representation to the first quantization representation. The left hand shows a state of three indistinguishable particles, and the right hand shows superposition between three states of three distinguishable particles symmetrized to fulfil the boson statistics.}
\label{fig:2ndto1st}
\end{figure}

While the mixed spin symmetry bases are defined in Eq.~\eqref{eq:basis_mixspin} in the first quantization representation, we adopt the second quantization representation for exact diagonalization. Therefore, we need to transform the second quantization representation of the Hamiltonians~\eqref{eq:H0_2nd}\eqref{eq:Hint_2nd} to the first quantization representation to see the spin symmetry. In the second quantization representation, we do not distinguish each particle and register only which energy level is occupied (see the left hand in Fig.~\ref{fig:2ndto1st}). On the other hand, in the first quantization representation each particle is labelled, and we register which energy level each particle is in in order of the labels. Considering bosons, each basis in the second quantization representation is equivalent to superposition of all the combinations of the labelling (see the right hand in Fig.~\ref{fig:2ndto1st}). For instance, replacing particles labelled by 1 and 2 does not change anything.
In this way, the bases in the second quantization representation can be converted into the bases in the first quantization representation, which give us the quantum number and the pseudo-spin state of each basis, i.e., each basis is described as $\ket{n_1,\sigma_1; n_2, \sigma_2; n_3, \sigma_3}$ for $n_1,n_2,n_3=0,1,\ldots,M-1$ and $\sigma_1,\sigma_2,\sigma_3=\downarrow,\uparrow$. 
Finally, we transform the pseudo-spin basis $\downarrow,\uparrow$ to the classified pseudo-spin basis $S_j$, $M_j$ by using the relations given in Eqs.~\eqref{eq:basis_symmetric}\eqref{eq:basis_mixspin}.

\section{Numerical results}
\label{sec:groundstate}

We construct the Hamiltonains~\eqref{eq:H0_2nd}\eqref{eq:Hint_2nd} for three particles and diagonalize them in a truncated space numerically. In this work, we fix $g_{\downarrow\downarrow}=g_{\uparrow\uparrow}=g$ to keep the symmetry between down and up spins. Moreover, to study competition between SOC and contact interaction, we turn on the anti-aligned pseudo-spin interaction $g_{\downarrow\uparrow}$ or the aligned pseudo-spin interaction $g$. The study of the two particle system has shown that these two cases give clear different properties in the ground state~\cite{Usui2020Spin}.
We use the trap energy $\hbar\omega$ as the energy unit and normalise the parameters with the trap energy $\hbar\omega$ and the trap length $\ell=\sqrt{\hbar/m\omega}$, for example the interaction strengths $g,g_{\downarrow\uparrow}$ are normalised with $\gamma\equiv\hbar\omega\ell$.
Also, we fix $k\ell=4$ and set the cutoff of the truncated space as $M=50$. We discuss the justification of the cutoff $M$ in Appendix~\ref{app:cutoff}, and our numerical code can be viewed in Ref.~\cite{Usui2023Code}. 





\begin{figure*}[t] 
\includegraphics[width=.8\linewidth]{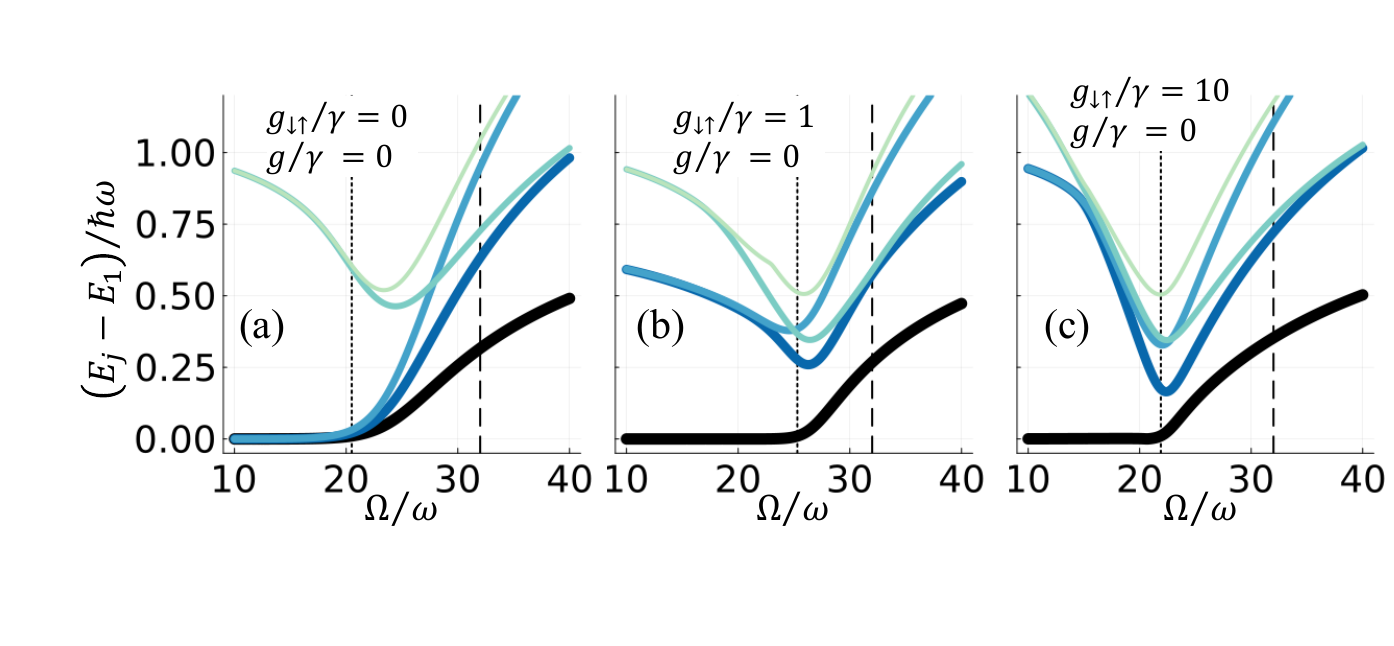}
\caption{Energy difference $E_j-E_0$ between the first to fifth excited states and the ground state for (a) the non-interacting case, (b) $g_{\downarrow\uparrow}/\gamma=1$, and (c) $g_{\downarrow\uparrow}/\gamma=10$. The dashed line is the critical point $\Omega_{\text{c}}^0/\omega=2 k^2\ell^2$ in free space, and the dotted line is when $E_1$ and $E_0$ start to deviate, $(E_1-E_0)/\hbar\omega>0.01$.}
\label{fig:spectrum_g12}
\end{figure*}

\subsection{Zero interactions $(g=g_{\downarrow\uparrow}=0)$}

It is known that SOC systems can have three different ground state phases~\cite{Li2012Quantum}: the stripe phase, the magnetised phase, and the single minimum phase. In the stripe phase, the ground state is a superposition of positive and negative momenta and thus displays an interference pattern. In the magnetised phase, the ground state acquires positive or negative momentum, and in the single minimum phase the spectrum only possess a single minimum, leading to zero momentum usually. Without contact interactions, only the stripe phase and the single minimum phase exist, and at the critical point some degeneracies are resolved. In free space the critical point between these two phases is given analytically by $ \Omega_{\text{c}}^0/\omega=2 k^2\ell^2$, and in trapped systems the density modulation modifies the critical value a lower value~\cite{Zhu2016Harmonically}. 

We plot the energy difference $E_j-E_0$ between some excited states and the ground state for the noninteracting case ($g=g_{\downarrow\uparrow}=0$) for changing the coherent coupling strength $\Omega$ and reconfirm that the degeneracy in the ground state is resolved off from $ \Omega_{\text{c}}^0$ (see the dashed line in Fig.~\ref{fig:spectrum_g12}(a)). Therefore, we define another critical point $\Omega_{\text{c}}$ for the trap system as a point where the energy difference $(E_1-E_0)/\hbar\omega$ between the ground state and the first excited state is larger than $10^{-2}$ (see the dotted line).
In the absence of coherent coupling, the ground states are four-fold degenerate, and their pseudo-spin states are aligned pseudo-spins such as (i) $\ket{\uparrow\uparrow\uparrow}$ and (ii) $\ket{\downarrow\downarrow\downarrow}$ and unaligned pseudo-spins such as (iii) $\ket{\uparrow\uparrow\downarrow}+\ket{\uparrow\downarrow\uparrow}+\ket{\downarrow\uparrow\uparrow}$ and (iv) $\ket{\uparrow\downarrow\downarrow}+\ket{\downarrow\uparrow\downarrow}+\ket{\downarrow\downarrow\uparrow}$.
Positive (negative) momentum is induced to down (up) spins due to SOC. The coherent coupling mixes these states, and in the strong coherent coupling limit an equally-weighted superposition of all pseudo-spin states obtains the lowest energy. 

By adding contact interactions between pseudo-spins, the above four degenerate ground states~(i-iv)
obtain different energy. We discuss the ground state in the case of repulsive anti-aligned interactions ($g_{\downarrow\uparrow}>0$ and $g=0$) first and in the case of repulsive aligned interactions ($g>0$ and $g_{\downarrow\uparrow}=0$) later.

\subsection{Anti-aligned interactions ($g_{\downarrow\uparrow}>0$ and $g=0$)}


We consider non-zero positive anti-aligned interaction $g_{\downarrow\uparrow}>0$, i.e. when the unaligned pseudo-spin components suffer from repulsive interaction. As a result, in the absence of the coherent coupling only the aligned pseudo-spin states $\ket{\downarrow\downarrow\downarrow}$, $\ket{\uparrow\uparrow\uparrow}$
obtain the lowest energy. 
We have computed the energy spectrum for relatively weak interaction $g_{\downarrow\uparrow}/\gamma=1$ and for strong interaction $g_{\downarrow\uparrow}/\gamma=10$. In the former (latter) case, the energy shift due to the contact interaction is less (more) than $\hbar\omega$. 
It is found that the degeneracy between the ground state and the first excited state is resolved at a different value of $\Omega$ from the noninteracting case, and it depends on the interaction strength $g_{\downarrow\uparrow}$ (see the dotted lines in Fig.~\ref{fig:spectrum_g12}(b,c)). 
Such deviation from the noninteracting case is also seen in the two-particle system but slightly smaller than in the three-particle system. This is because in the three-particle system the contact interaction energy is larger than in the two-particle system, and the maximum energy shift due to contact interaction is $\hbar\omega$ for two particles and $3\hbar\omega$ for three particles.
For $\Omega/\omega\simeq 40$, the energy spectrum for different $g_{\downarrow\uparrow}$ look similar to each other because the coherent coupling is so strong that the contact interaction contribution to the energy is negligible. 

\begin{figure*}[t] 
\includegraphics[width=.99\linewidth]{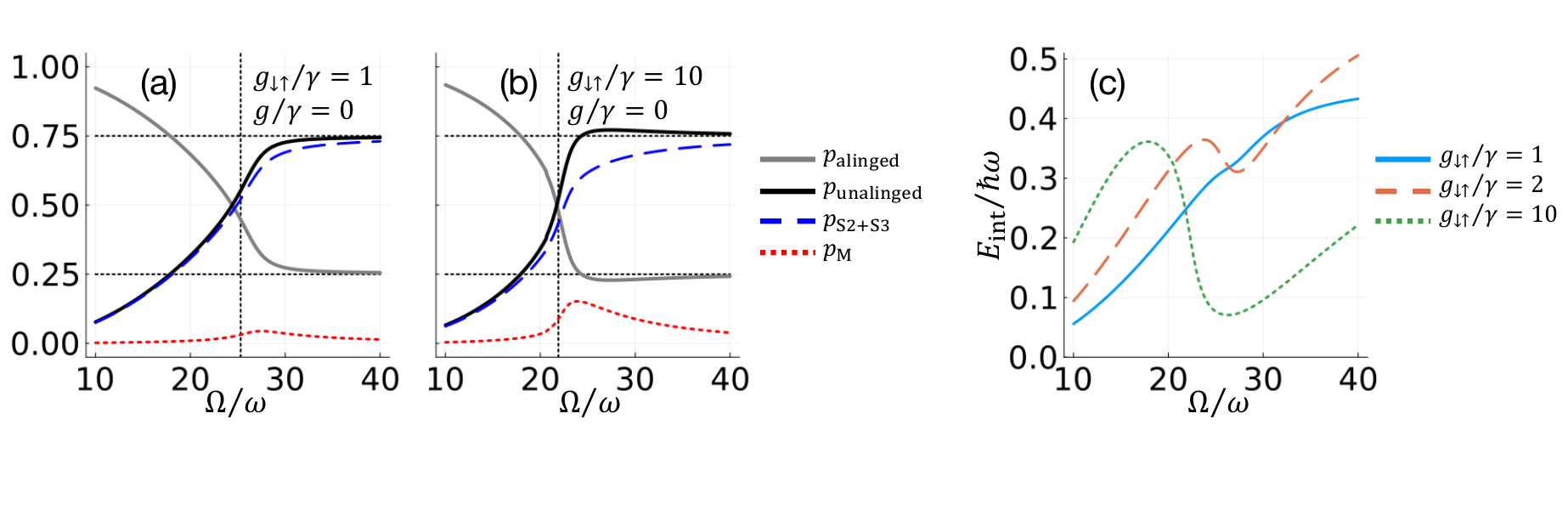}
\caption{(a,b) Pseudo-spin population $p_{\text{aligned}}$, $p_{\text{unaligned}}$ in the ground state for weak or strong contact interactions in the three-particle system. The dashed and dotted lines represent the population $p_{S_2+S_3}$ of $\ket{S_2}$ and $\ket{S_3}$ and the population $p_{M}$ of the mixed spin symmetry states, respectively. Note that $p_{\text{unaligned}}=p_{S_2+S_3}+p_M$. The dotted vertical lines represent $\Omega_c$. (c) Interaction energy for different $g_{\downarrow\uparrow}$.}
\label{fig:spinpop_g_antialinged}
\end{figure*}

\subsubsection{Pseudo-spin population}

We study the pseudo-spin component in the ground state. 
Here, we categorise the pseudo-spin population of the ground state into the aligned pseudo-spin components and the unaligned pseudo-spin components, i.e. the population $p_{\text{aligned}}$ of $\ket{\downarrow\downarrow\downarrow}$ and $\ket{\uparrow\uparrow\uparrow}$ and the population $p_{\text{unaligned}}$ of the rest of the pseudo-spin basis, therefore $p_{\text{aligned}}+p_{\text{unaligned}}=1$. Note that the populations of $\ket{\downarrow\downarrow\downarrow}$ and $\ket{\uparrow\uparrow\uparrow}$ are the same due to the symmetry of the contact interactions we set ($g_{\downarrow\downarrow}=g_{\uparrow\uparrow}$). 
We have plotted these pseudo-spin components $p_{\text{aligned}}$, $p_{\text{unaligned}}$ as a function of coherent coupling strength $\Omega$ (see Fig.~\ref{fig:spinpop_g_antialinged}(a,b)). 
For weak coherent coupling $\Omega$, the aligned pseudo-spin population $p_{\text{aligned}}$ is dominant because of repulsive anti-aligned interaction. By increasing $\Omega$, the unaligned pseudo-spin component $p_{\text{unaligned}}$ also grows. Eventually, the population $p_{\text{aligned}}$ is approaching $3/4$ while $p_{\text{unaligned}}$ is approaching $1/4$, because the lowest energy state in the limit $\Omega\to\infty$ is 
\begin{align}
    (\ket{\uparrow}-\ket{\downarrow})^{\otimes 3}
    &=
    \ket{\uparrow\uparrow\uparrow} -\ket{\uparrow\uparrow\downarrow} - \ket{\uparrow\downarrow\uparrow} - \ket{\downarrow\uparrow\uparrow}
    \nonumber
    \\
    &\quad +
    \ket{\uparrow\downarrow\downarrow} +\ket{\downarrow\uparrow\downarrow} + \ket{\downarrow\downarrow\uparrow} - \ket{\downarrow\downarrow\downarrow}.
    \nonumber
\end{align}
In the intermediate regime, these populations $p_{\text{aligned}}$, $p_{\text{unaligned}}$ grow differently depending on $g_{\downarrow\uparrow}$. While for the weak interaction case~(a) in Fig.~\ref{fig:spinpop_g_antialinged} both populations increase monotonically, for the strong interaction case~(b) in Fig.~\ref{fig:spinpop_g_antialinged} the population $p_{\text{aligned}}$ ($p_{\text{unaligned}}$) reaches an extreme value and overcomes $1/4$ $(3/4)$, i.e. $p_{\text{aligned}}$ is concave, and $p_{\text{unaligned}}$ is convex.  

\begin{figure}[b] 
\includegraphics[width=.8\linewidth]{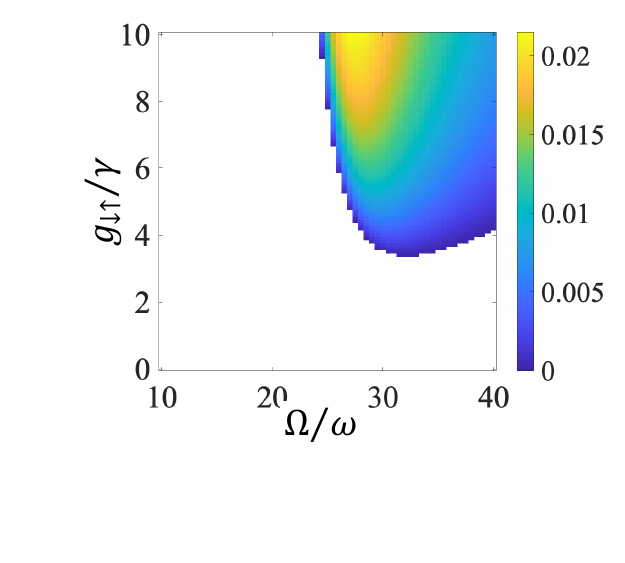}
\caption{
Excess of $p_{\text{unaligned}}$ over the population at the limit of $\Omega\to\infty$ as function of $\Omega/\omega$ and $g_{\downarrow\uparrow}/\gamma$, i.e. $p_{\text{unaligned}}-3/4$. Negative values are not shown for clear presentation. 
}
\label{fig:spinpop_2d}
\end{figure}

Furthermore, we have computed the population $p_{\text{M}1}, p_{\text{M}2}, p_{\text{M}3}, p_{\text{M}4}$ of the mixed spin symmetric basis~\eqref{eq:basis_mixspin}. 
Since these populations are the same due to our setting $g_{\downarrow\downarrow}=g_{\uparrow\uparrow}$, we have plotted the sum  $p_{\text{M}}=p_{\text{M}1}+p_{\text{M}2}+p_{\text{M}3}+p_{\text{M}4}$ (see the dotted lines in Fig.~\ref{fig:spinpop_g_antialinged}(a,b)). 
We have found that a large amount of the mixed spin symmetric states appear in the ground state for the strong interaction case (b) in Fig.~\ref{fig:spinpop_g_antialinged}. Considering $p_{\text{unaligned}}=p_{\text{S}2+\text{S}3}+p_{\text{M}}$ with $p_{\text{S}2+\text{S}3}$ the population of $\ket{S_2}$ and $\ket{S_3}$, this leads to the non-monotonic growth of $p_{\text{aligned}}$, $p_{\text{unaligned}}$. This is also seen in the two-particle system, and in that case it is originated from the emergence of the anti-symmetric pseudo-spin states~\cite{Usui2020Spin}. The interaction energy increases for increasing the contact interaction strength in general, but the anti-symmetric pseudo-spin states do not feel contact interactions. Thus, the appearance of the anti-symmetric pseudo-spin states for strong contact interactions suppresses the interaction energy.
In three-particle systems, the mixed spin symmetric states is mixture of spin symmetric states and spin antisymmetric states, and this anti-symmetry reduces the interaction energy.
We have calculated the interaction energy $E_{\text{int}}=\langle\hat{H}_{\text{int}}\rangle$ and plotted it as a function of $\Omega$, which displays a dent for strong interaction (see Fig.~\ref{fig:spinpop_g_antialinged}(c)).   
Finally, we have inspected the non-monotonic growth of $p_{\text{unaligned}}$ for a wide range of interaction $g_{\downarrow\uparrow}$ (see Fig.~\ref{fig:spinpop_2d}) and confirmed that the excess over $3/4$ is larger for stronger $g_{\downarrow\uparrow}$. 

\subsubsection{Pair correlation}

\begin{figure*}[tb]
\centering
\includegraphics[width=.99\linewidth]{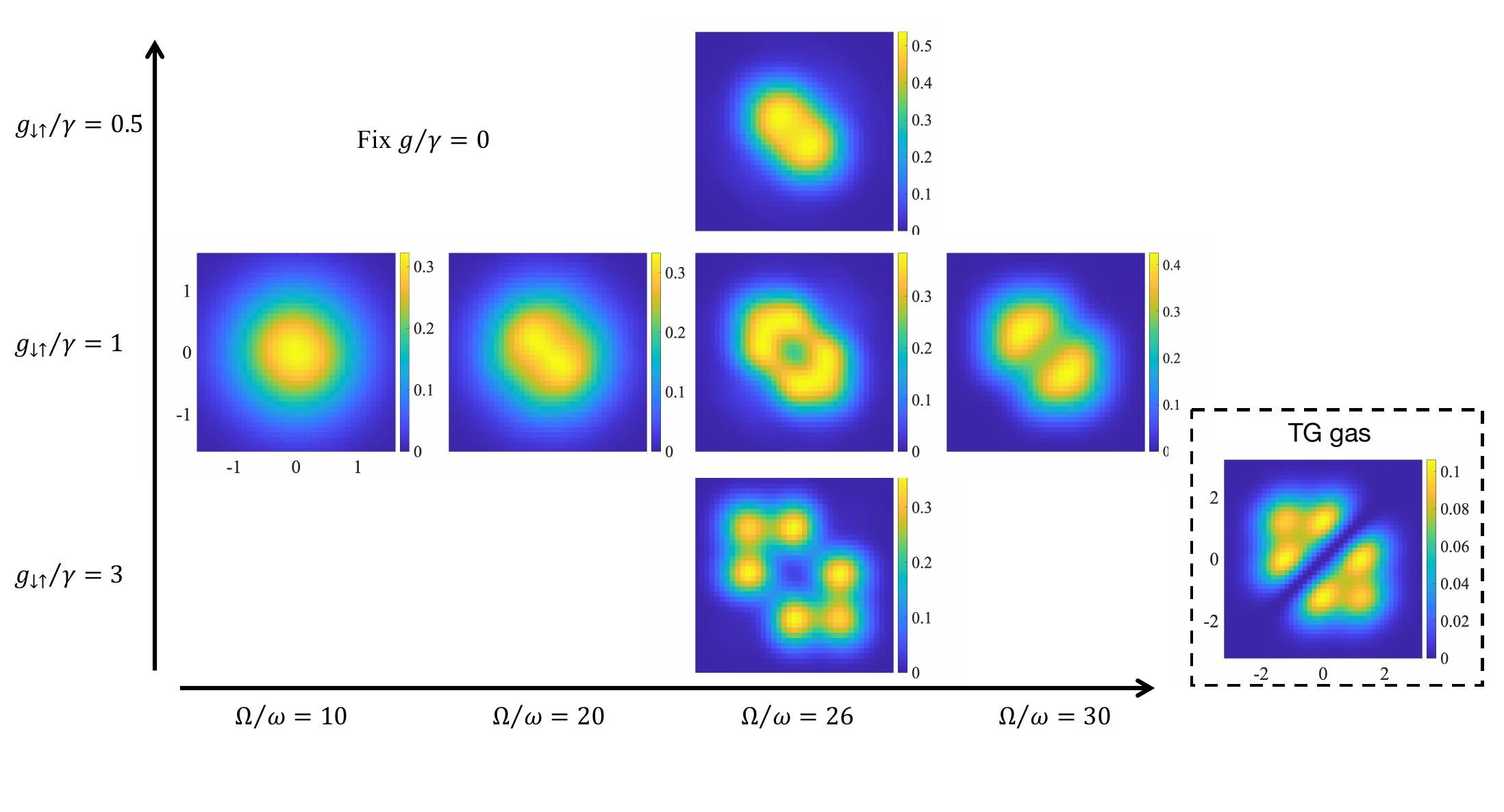}
\caption{Pair correlation of the ground state when changing $g_{\downarrow\uparrow}$ and $\Omega$ while fixing $g=0$. The right panel displays the pair correlation of TG gas, and notice that the scale of the right panel is different from the other panels.}
\label{fig:pair_g12}
\end{figure*}

So far we have shown that properties of the ground state in three particle systems are similar to two particle systems regardless of the difference of the spin symmetry. In this section, we discuss the spatial structure of the ground state and show that the interplay between SOC and contact interactions brings distinct features from the two-particle system.

The pair correlation of the ground state $\Psi_{\text{GS}}(x_1,x_2,x_3)$ is defined as
\begin{align}
    \rho_2(x,y)
    &=
    \int dx_3 \left| \Psi_{\text{GS}}(x,y,x_3) \right|^2,
\end{align}
where two of the spatial degrees are kept and the other is integrated out. In the second quantisation representation it is given by
\begin{align}
    \rho_2(x,y)
    &=
    \frac{1}{N(N-1)}
    \sum_{i,j,p,q=1}^{2M}
    \mel{\Psi_{\text{GS}}}{\hat{a}_i^{\dagger} \hat{a}_p^{\dagger} \hat{a}_j \hat{a}_q}{\Psi_{\text{GS}}}
    \nonumber\\
    &\quad\quad\quad\quad\quad\quad
    \phi_i^*(x)
    \phi_p^*(y)
    \phi_j(x)
    \phi_q(y)
\end{align}
with $\phi_i(x)$ the $i$th eigenstate of the harmonic oscillator and $N=3$ particle number~\cite{Mujal2017Quantum}.
The pair correlation of non-interacting particles is given by a two-dimensional Gaussian function, $\rho_2(x,y)=|\phi_0(x)|^2|\phi_0(y)|^2$. In the limit of strong interactions $g_{\downarrow\uparrow},g\to\infty$, the system corresponds to Tonks-Girardeau gas (TG) gas and behaves as (spinless) fermions~\cite{Tonks1936Complete,Lieb1963Exact,Garcia2014Quantum}, and particularly the density profile corresponds to that of the free fermions, $|\Psi_{\text{TG}}(x_1,x_2,x_3)|=|\Psi_{\text{F}}(x_1,x_2,x_3)|$. Therefore, the pair correlation in the limit can be calculated analytically by using the wavefunction of three free fermions trapped in the harmonic potential, given by
\begin{align}
    \Psi_{\text{F}}(x_1,x_2,x_3)
    &=
    \frac{1}{\sqrt{3!}}
    \begin{vmatrix}
    \phi_0(x_1) & \phi_1(x_1) & \phi_2(x_1) \\
    \phi_0(x_2) & \phi_1(x_2) & \phi_2(x_2) \\
    \phi_0(x_3) & \phi_1(x_3) & \phi_2(x_3)
    \end{vmatrix}
    , 
\end{align} 
and is plotted for reference in the right panel in Fig.~\ref{fig:pair_g12}. 

We study the pair correlations, and first let us focus on $g_{\downarrow\uparrow}/\gamma=1$ and change the coherent coupling strength $\Omega$ (see the middle panels at $g_{\downarrow\uparrow}=1$ in Fig.~\ref{fig:pair_g12}). For relatively small coherent coupling $\Omega/\omega=10$, the pair correlation is close to a Gaussian distribution, and the contact interaction affects the pseudo-spin populations but does not change the spatial structure of the ground state from the free particle case. By increasing the coherent coupling strength ($\Omega/\omega=20$), the shape is squished along $x=y$. At $\Omega/\omega=26$, 
a dent emerges at the centre ($x=y=0$). 
For increasing $\Omega$ more, a slit at $x=y$ becomes deeper, and two bumps remain. This slit is originated solely from the contact interaction, and the pair correlation remains in the same shape even for larger $\Omega$. It is consistent to the pseudo-spin populations $p_{\text{aligned}}$, $p_{\text{unaligned}}$, which reach plateau when $\Omega/\omega\simeq30$. 


To dig into the interplay between the contact interaction and the SOC, we fix the coherent coupling strength to $\Omega/\omega=26$ and change the contact interaction strength (see the panels at $\Omega=26$ in Fig.~\ref{fig:pair_g12}). 
By decreasing the contact interaction strength, the dent seen for $g_{\downarrow\uparrow}/\gamma=1$ becomes invisible. On the other hand, for increasing $g$ the pair correlation shows three bumps in $x>y$ and in $x<y$. This is a qualitatively similar feature to the TG gas other than the size of these bumps. It is interesting that such similarly to TG gas is seen even for relatively small interaction such as $g_{\downarrow\uparrow}/\gamma=3$. For larger $g_{\downarrow\uparrow}$, the same structure remains. 
Also, it is worth noting that there is 
no interaction between the same pseudo-spins. 
In the two-particle system, the counterpart of the pair correlation is density distribution $\rho_2(x_1,x_2)=|\Psi_{\text{GS}}(x_1,x_2)|^2$, and it was investigated in our previous work~\cite{Usui2020Spin}. However, such strong interaction feature enhanced by SOC is not observed.



\begin{figure}[t] 
\centering
\includegraphics[width=.99\linewidth]{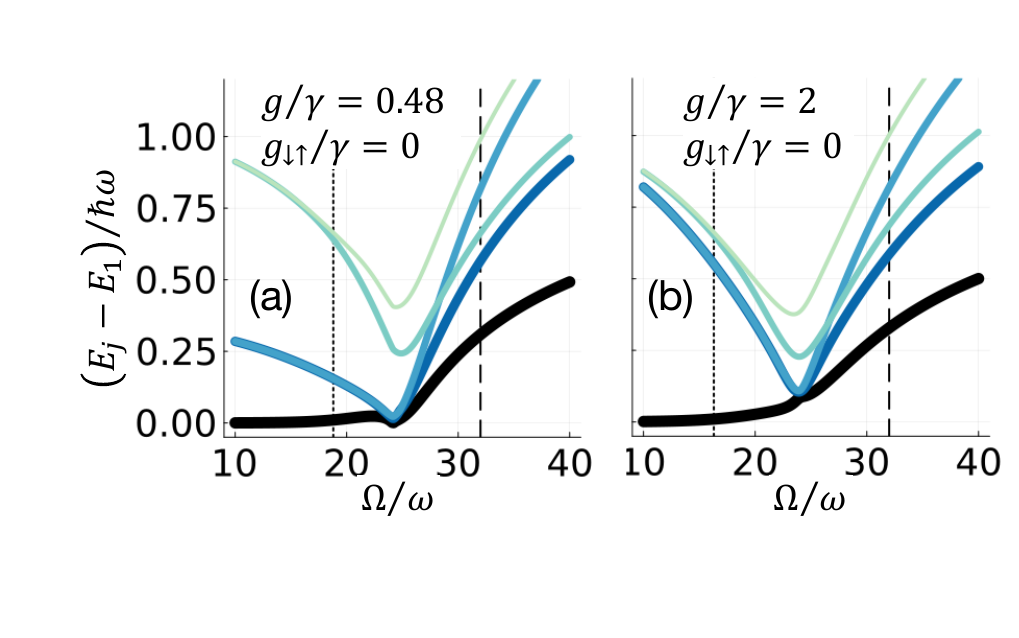}
\caption{
Energy difference $E_j-E_0$ between some excited states and the ground state for $g/\gamma=0.48$ (a) and $g/\gamma=2$ (b). The dashed line is the critical point $\Omega_{\text{c}}^0/\omega=2 k^2\ell^2$ in free space, and the dotted line is when $E_1$ and $E_0$ start to deviate, $(E_1-E_0)/\hbar\omega>0.01$.
}
\label{fig:panels_g}
\end{figure}

\subsection{Aligned interaction ($g>0$ and $g_{\downarrow\uparrow}=0$)}




Now we turn off anti-aligned interaction $g_{\downarrow\uparrow}=0$ and set non-zero aligned interaction strength $g>0$. 
Any pseudo-spin configuration of three particle states has the aligned pseudo-spin components and is affected by the aligned interaction. For $\Omega/\omega=0$, the two unaligned pseudo-spin states (the states~(iii,iv) mentioned in Sec.~\ref{sec:groundstate})
obtain the lowest energy. 
For finite $\Omega$, the coherent coupling mixes those states and the two pseudo-spin aligned components $\ket{\downarrow\downarrow\downarrow}$, $\ket{\uparrow\uparrow\uparrow}$.
In the limit $\Omega\to\infty$, the contact interaction contribution is not visible, and the ground state approaches that of no interactions. 

Figure~\ref{fig:panels_g} reveals the energy difference $E_j-E_0$ in the intermediate regime of $\Omega$ for two different interaction strengths $g/\gamma=0.48, 2$. It is similar to the anti-aligned interaction case $(g_{\downarrow\uparrow}>0)$ that the critical points $\Omega_{\text{c}}$ vary in different interaction strengths. 
One different feature is that the avoided crossing between the ground state and the first excited states is seen for $g/\gamma=0.48$ when $\Omega/\omega\simeq25$ but not for $g/\gamma=2$.
This is caused by the interplay between the contact interaction and the coherent coupling that have attempt to shift energy in the opposite directions~\cite{Guan2014energy}. By increasing $g$, the excited states are pushed up, and accordingly the avoided crossing is shifted up and shades out eventually. The existence of this avoided crossing affects the pseudo-spin population as see below. 

\subsubsection{Pseudo-spin population}

\begin{figure}[t] 
\includegraphics[width=.8\linewidth]{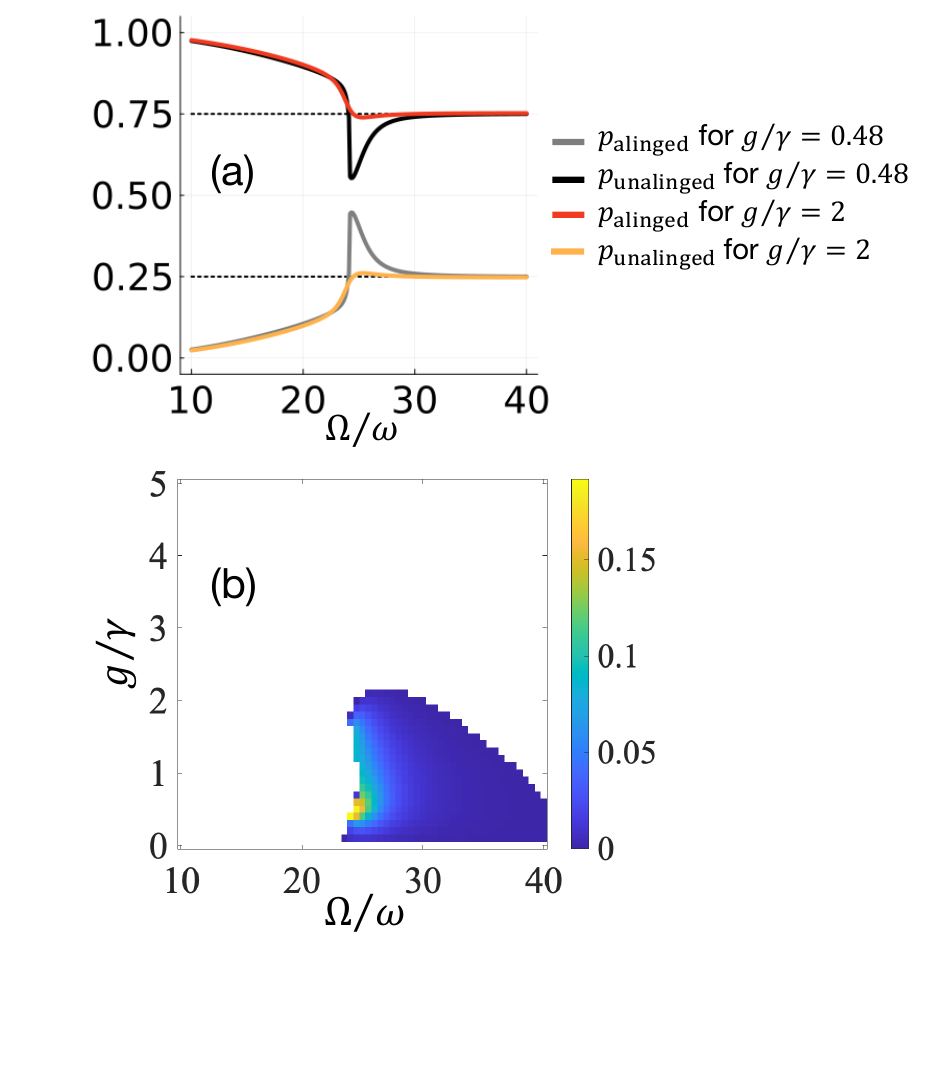}
\caption{
(a) Pseudo-spin population $p_{\text{aligned}}$, $p_{\text{unaligned}}$ in the ground state for $g/\gamma=0.48$ and $g/\gamma=2$.
(b) Excess of $p_{\text{aligned}}$ over the population at the limit of $\Omega\to\infty$ as function of $\Omega/\omega$ and $g/\gamma$, i.e. $p_{\text{aligned}}-1/4$. Negative values are not shown for clear presentation. 
}
\label{fig:spinpop_g_alinged}
\end{figure}

As shown, in the anti-aligned interaction case ($g_{\downarrow\uparrow}>0$), the pseudo-spin populations $p_{\text{aligned}}$, $p_{\text{unaligned}}$ become larger than the values in the limit $\Omega\to\infty$ for larger interaction strength. In the aligned interaction case ($g>0$), the aligned pseudo-spin population $p_{\text{aligned}}$ jumps up and the unaligned pseudo-spin population $p_{\text{unaligned}}$ drops down for relatively weak interaction (see Fig.~\ref{fig:spinpop_g_alinged}(a)).
The emergence of this sharp change matches the appearance of the avoided crossing, and for large $g$ these extreme values of $p_{\text{aligned}}$ and $p_{\text{unaligned}}$ disappear (see Fig.~\ref{fig:spinpop_g_alinged}(b)). 
This is contrasted with the anti-aligned interaction case, where the pseudo-spin populations $p_{\text{aligned}}$ and $p_{\text{unaligned}}$ obtain the extreme values when $\Omega=\Omega_c$, i.e. when the degeneracy between the first excited state and the ground state is resolved. 
This non-monotonic behaviour survives only for small values of $g$ compared to the anti-aligned interaction case because the emergence of aligned pseudo-spin states does not reduce the interaction energy.
We note that such population jump has been also observed in the two-particle system~\cite{Usui2020Spin}. 


\subsubsection{Pair correlation}



\begin{figure*}[t]
\centering
\includegraphics[width=.99\linewidth]{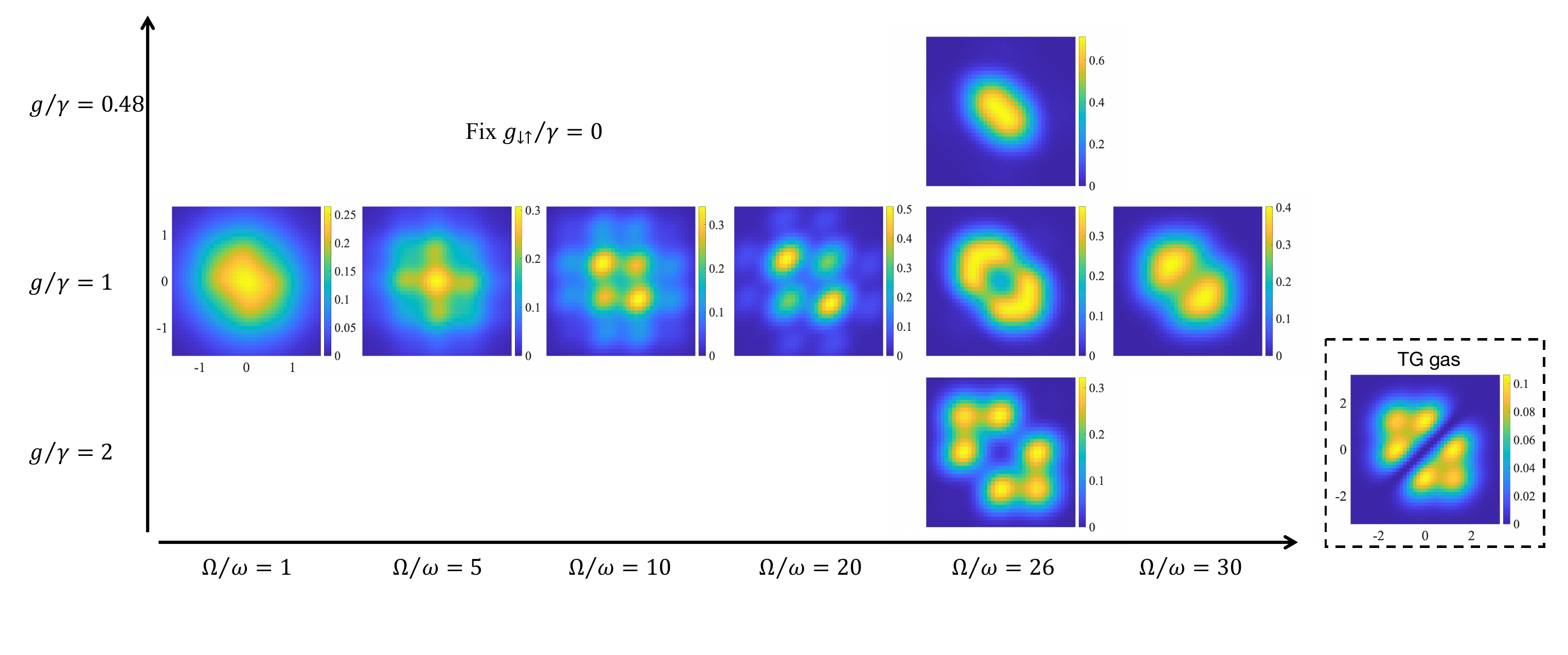}
\caption{Pair correlation of the ground state when changing for $g$ and $\Omega$ while fixing $g_{\downarrow\uparrow}=0$. The right panel displays the pair correlation of TG gas, and notice that the scale of the right panel is different from the other panels.}
\label{fig:pair_g}
\end{figure*}

We have computed the pair correlations for different $g$ and $\Omega$. First, let us focus on $g/\gamma=1$ and change $\Omega$ (see the middle panels at $g=1$ in Fig.~\ref{fig:pair_g}). 
Overall, the pair correlation behaves similar to the anti-aligned interaction case but is more sensitive to $\Omega$. Even for relatively small coupling, the pair correlation starts deviating from Gaussian distribution. For $\Omega/\omega=5$, one bump appears at the centre ($x=y=0$), and for increasing $\Omega$ four bumps come out. 
At $\Omega/\omega=26$, where the populations $p_{\text{aligned}}$, $p_{\text{unaligned}}$ approach the extreme values as shown in Fig.~\ref{fig:panels_g}, a dent emerges at the centre. 
For increasing $\Omega$ more, a slit at $x=y$ caused by the contact interaction appears, and the effect of SOC vanishes. 

Now we fix the coherent coupling strength to $\Omega/\omega=26$ and change the contact interaction strength to $g/\gamma=0.48,2$. 
For $g/\gamma=0.48$, the dent at $x=y=0$ vanishes, and for $g/\gamma=2$ three bumps emerge in $x>y$ and in $x<y$ respectively, which is similar to the anti-aligned interaction case. The aligned interaction and the anti-aligned interaction give similar effects to the structure of the pair correlations even though they affect the energy spectrum and the pseudo-spin population differently.


\section{Conclusions} \label{sec:conclusions}

We have studied the system of three bosons trapped in a harmonic potential with two internal states in the presence of SOC and investigated the ground state and the energy spectrum by implementing exact diagonalization. 
Particularly, we have focused on how the mixed spin symmetric states contribute to the ground state and have found similarity and difference.
We have found that the interplay between the anti-aligned contact interaction and SOC increases the population of the unaligned pseudo-spin components in the ground state. 
This is due to the emergence of the mixed spin symmetric states, which can be partially anti-symmetric. This anti-symmetry compensates for the rise of the interaction energy.
Moreover, with the aligned interactions, the avoided crossing between the ground state and the first excited state is observed only for small interaction, and this causes shape changes in the pseudo-spin populations. However, the emergence of aligned pseudo-spin states does not contribute to the reduction of the interaction energy, and therefore it decays out for strong interaction.
This is similar to the two-particle system. 
In addition, we have found that the pair correlation of the ground state has a structure similar to the TG gas even for relatively small contact interactions. This strong interaction effect is not seen in two-particle systems and is interesting because there are only aligned or anti-aligned interactions.

A natural question is whether the same behaviour is seen in the more-particle systems. Our results cannot guarantee but imply that this would be the case for anti-aligned interactions. In these cases, the mixed spin symmetry affects the pseudo-spin populations, and more-particle systems also have the same spin symmetry. The mixed spin symmetric states may act in the same way. 
Furthermore, the results we have shown may change by inducing different types of interactions. For example, here we have studied only repulsive contact interactions between two particles where anti-symmetric pseudo-spin states do not feel the interaction. This property would be different if we were also to consider three-body interactions~\cite{Hammer2013Colloquium}.

Additionally, we have focussed on one dimension, but it is possible to induce SOC in two dimensions~\cite{Mujal2020Spin,Wu2016Realization}. There are some differences in SOC in one dimension and two dimensions. 
In two dimensions, there are the Rashba type and the Dresselhaus type of SOC~\cite{Schillaci2015Energy}, and they are given by $\left(\hat{p}_x\hat{\sigma}_x\pm\hat{p}_y\hat{\sigma}_y\right) \hbar k/m$, respectively. By inducing both types with equal weight, the SOC term is given by $2\hat{p}_x\hat{\sigma}_x \hbar k/m$, which is essentially the same term considered in one dimension after rotating the frame. Therefore, the results we have shown would be observed in two dimensions when such type of SOC is applied~\cite{Lin2011Spin}. Otherwise, unique effects to two dimensions could be seen~\cite{Mujal2020Spin,Schillaci2015Energy}.
Also, we have compared our system with a TG gas in finding qualitative similarity of the SOC system to the strongly interacting system. TG gas is unique to one dimension and does not exist in higher dimensions. It is an interesting question whether such enhancement of strong interaction effects is still observed in higher dimensions.

The mapping between Dicke model and SOC systems with contact interactions was proposed~\cite{Hamner2014Dicke,Lian2013Orbit,Huang2015Spin}, assuming that the pseudo-spin space of the SOC system is only symmetric. However, as shown already, there is significant amount of mixed spin symmetric components even in the ground state for some parameter regime. Therefore, the mapping between these two models is limited. 

In this work, we focus on the low energy states, but it is possible to study excited states and quench dynamics in a wide parameter range with our numerical method. As future work, it would be interesting to study transport via SOC~\cite{Chen2018Inverse} in the presence of strong interaction and utilise SOC, which gives different directions of momentum to different pseudo-spins, to create spatial entanglement~\cite{Fadel2018Spatial,Philipp2018Spatially,Karsten2018Entanglement}. 


\section{acknowledgement}
Discussions with Thomas Busch, Karol Gietka, Yongping Zhang, Peter Engels, Pere Mujal, and Ian B. Spielman are gratefully appreciated.
We acknowledge funding from Grant No.~PID2020-114626GB-I00 by MCIN/AEI/10.13039/5011 00011033 and ``Unit of Excellence Mar\'ia de Maeztu 2020-2023'' award to the Institute of Cosmos Sciences, Grant CEX2019-000918-M funded by MCIN/AEI/10.13039/501100011033. 
We also acknowledge financial support from the Generalitat de Catalunya (Grant 2021SGR01095).
A.U acknowledges further support from the Agencia Estatal de Investigaci\'{o}n and the Ministerio de Ciencia e Innovaci\'{o}n.
A.R.-F. acknowledges funding from MIU through Grant No. FPU20/06174.
\appendix

\section{Justification of cut off $M$} \label{app:cutoff}

In Sec.~\ref{sec:groundstate}, we set the SOC strength $k\ell=4$ and the cut off $M=50$. Here, we show that the cut off $M$ is large enough. 
As discussed in Sec.~\ref{sec:2ndquantization}, the cut off $M$ is the number of the eigenstates of the harmonic oscillator we take. The SOC couples these eigenstates such that the momentum of up spin or down spin is boosted. Although larger SOC needs larger $M$, the deviation from the results when $k=0$ is negligible if $M$ is large enough. 

\begin{figure}[tb]
\centering
\includegraphics[width=.9\linewidth]{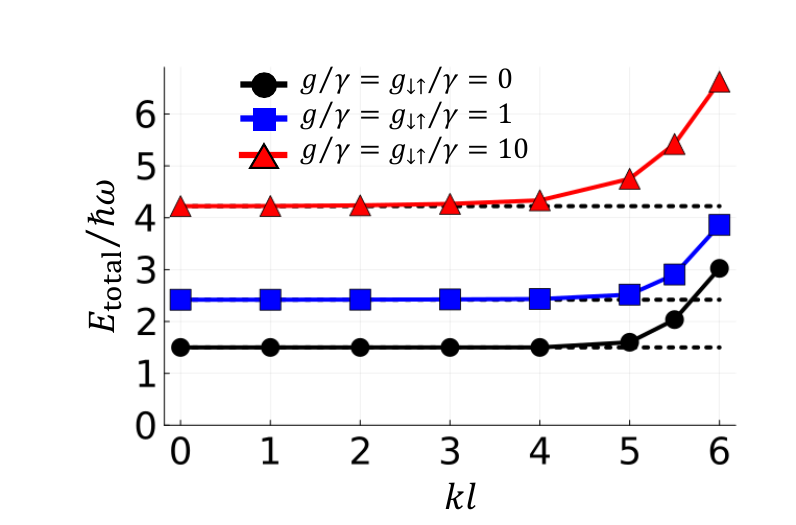}
\caption{Total energy for different interaction strengths $g=g_{\downarrow\uparrow}$, where $\Omega=0$ and $M=50$ are set. The dotted lines show the analytical results of the energy for $k=0$.}
\label{fig:ene_error}
\end{figure}

Without coherent coupling $\Omega$, the Hamiltonian~\eqref{eq:Htotal} can be diagonalised in the basis of $\sigma_z$, and the lowest energy can be obtained anaalytically and is the same as when $k=0$. For instance, the kinetic term of the single particle Hamiltonian~\eqref{eq:H0} can be written as $\sum_{j=1}^{N}(p_j-\hbar k\sigma_z)^2/2m$. The SOC acts as a moving frame, and the lowest energy does not change. Thus, by looking at the lowest energy for different $k$, we see whether the cut off $M$ is large enough. Figure~\ref{fig:ene_error} shows the energy computed with $M=50$ as function of $k$. For large $k$, the energy obtained numerically deviates from the energy with $k=0$. 
For strong interactions, the deviation is worse because strong interactions also couple highly excited eigenstates. 
For $k\ell=4$, the error is small and at worst about $2.7\%$ for $g/\gamma=g_{\downarrow\uparrow}/\gamma=10$, and we have taken $k\ell=4$ in the main text. 

\bibliography{bibliografia}

\end{document}